\documentclass{sig-alternate}
\usepackage{amssymb}
\usepackage{amsmath}
\usepackage{multirow}
\usepackage{enumerate}
\usepackage{float}
\usepackage{graphicx}
\usepackage{algorithm}
\usepackage{algpseudocode}
\usepackage{multirow}

\begin{document}
\title{An Inference Attack Model for Flow Table Capacity and Usage: Exploiting the Vulnerability of Flow Table Overflow in Software-Defined Network}
\author{
{Junyuan Leng$^{*}$,}
{Yadong Zhou$^{\dag}$,}
{Junjie Zhang$^{\ddag}$,}
{Chengchen Hu$^{\S}$}\\
\affaddr{$^{*}${}$^{\dag}${}$^{\S}$MOE Key Lab for Intelligent Networks and Network Security, Xi'an Jiaotong University}\\
\affaddr{$^{\ddag}$Department of Computer Science and Engineering, Wright State University}\\
\affaddr{\{$^{*}$jyleng, $^{\dag}$ydzhou\}@sei.xjtu.edu.cn $^{\ddag}$junjie.zhang@wright.edu $^{\S}$huc@ieee.org}
}
\maketitle
\begin{abstract}
As the most competitive solution for next-generation network, software-defined network (SDN) and its dominant implementation OpenFlow, are attracting more and more interests. But besides convenience and flexibility, SDN/OpenFlow also introduces new kinds of limitations and security issues. Of these limitations, the most obvious and maybe the most neglected one, is the flow table capacity of SDN/OpenFlow switches.

In this paper, we proposed a novel inference attack targeting at SDN/OpenFlow network, which is motivated by the limited flow table capacities of SDN/OpenFlow switches and the following measurable network performance decrease resulting from frequent interactions between data plane and control plane when the flow table is full. To our best knowledge, this is the first proposed inference attack model of this kind for SDN/OpenFlow. We also implemented an inference attack framework according to our model and examined its efficiency and accuracy. The simulation results demonstrate that our framework can infer the network parameters(flow table capacity and flow table usage) with an accuracy of 80\% or higher. These findings give us a deeper understanding of SDN/OpenFlow limitations and serve as guidelines to future improvements of SDN/OpenFlow.
\end{abstract}


\section{Introduction}
By decoupling the control plane from the data plane, Software-Defined Network (SDN) makes programmability a built-in feature for networks, thereby introducing automaticity and flexibility to the networking management. SDN has therefore been foreseen as the key technology that enables the next generation of networking paradigm. Despite its promise, one of the most significant barriers towards SDN's wide practical deployment resides in overwhelming security concerns. Therefore, proactively detecting, quantifying, and mitigating its security vulnerabilities becomes of fundamental importance. 

In spite of its novelty, SDN indeed reuses various design and implementation elements ranging from architectures and protocols to systems from traditional network. It is not surprising that SDN inheres the vulnerabilities intrinsic to these elements. For example, similar to any networked service, secure channels between controllers and switches might be disrupted by DDoS attacks; like firewall rules, the flow entries may also conflict with each other, leaking unwanted traffic; malicious arp spoofing generated by attackers may poison the controller MAC table, disturbing the normal topology information gathering and packet forwarding; untrusted applications may instrument SDN controller to perform malicious behaviors without proper access control, which is one of the design objectives for modern operating systems. In response, existing research in the context of SDN security mainly focuses on detecting and mitigating these vulnerabilities. For example, ~\cite{OpenFlow Vulnerability Assessment} evaluates man-in-the-middle attacks that target at SDN/OpenFlow secure channels; FortNOX~\cite{FortNOX} brings security enforcement module into NOX~\cite{NOX} and enables real-time flow entry conflict check; VeriFlow~\cite{VeriFlow} detects network-wide invariant violations by acting as a transparent layer between control plane and data plane.

In this paper, we introduce a novel SDN vulnerability. The novelty of this vulnerability stems from the feedback-loop nature of SDN, a fundamental difference compared with traditional networks. 

Specifically, most commercial SDN/OpenFlow switches have limited flow table capacities, ranging from hundreds to thousands~\cite{PAST}. Such capacity is usually insufficient to handle millions of flows that are typical for enterprise and data center networks~\cite{Network traffic characteristics of data centers in the wild}. Nevertheless, the flow table capacity was just considered as a potential bottleneck of resource consuming attacks in the past, motivating researches on flow caching systems like \cite{Flow caching for high entropy packet fields}, \cite{CAB} and \cite{CacheFlow}. But according to our analysis, the flow table capacity can lead to inference attack and privacy leakage under certain circumstances.

As a consequence of flow table overflow, the SDN controller needs to dynamically maintain the flow table by inserting and deleting flow entries. The maintaining process typically include packet information transferring, routing rule calculation and flow entry deployment, which leads to measurable network performance decrease.

Particularly, once the flow table is full, extra interactions between controller and switch are needed to remove certain existing flow entries to make room for newly generated flow entries,  resulting in further network performance decrease. An attacker can therefore leverage the perceived performance change to deduce the internal state of the SDN. To be more specific, we consider the scenario that an attacker resides in a network that is managed by a SDN. The attacker can then actively generate network traffic, triggering the interactions between the controller and switch with respect to flow entry insertion and deletion. The attacker can then measure the change of the network performance to estimate the internal state of the SDN including the flow table capacity and flow table usage. We have designed innovative algorithms to exploit this vulnerability and quantify their effectiveness on exploiting this vulnerability based on extensive evaluation.

To summarize, in this paper we made the following contributions:
\begin{itemize}
\item We have identified a novel vulnerability introduced by the limited flow table capacities of SDN/OpenFlow switches and formalized that threat.

\item We have designed effective algorithms that can successfully exploit this vulnerability to accurately infer the internal states of the SDN network including flow table capacity and flow table usage. 

\item We have performed extensive evaluation to quantify the effectiveness of proposed algorithms. The experimental results have demonstrated that the discovered vulnerability indeed leads to significant security concerns: our algorithm can infer the network parameters with an accuracy of 80\% or higher across various network settings.
\end{itemize}
The rest of this paper is organized as follows. Section \ref{chap:background} gives an overview of some background information. Section \ref{chap:problem-statement} gives an overall statement of the inference attack problem. Section \ref{chap:fifo-inference-algorithm} and \ref{chap:lru-inference-algorithm} give detailed inference algorithms targeting at FIFO and LRU replacement algorithms respectively. Section \ref{chap:evaluation} gives a detailed evaluation of the simulation results. Section \ref{chap:discussion} is a brief discussion about our findings and future research. Finally, section \ref{chap:conclusion} concludes this paper.
\section{Background}\label{chap:background}
\subsection{Software-Defined Network}
Software Defined Network (SDN) is a competitive solution for next-generation network. SDN offers network programmability by separating the control plane from the data plane. Network functions like routing calculation and link discovery are extracted from switches (data plane) and implemented by centralized controllers (control plane). OpenFlow~\cite{OpenFlow Introduction} is the most prominent SDN implementation.

In a SDN network, the controller gathers network topology information and makes high-level routing decisions while the switches only perform the functionality of packet forwarding according to routing rules assigned by the controller. The dedicated link connecting controller and switch is called secure channel. Controller and switch communicates via secure channel using the OpenFlow protocol. Controller also exposes network control APIs or north-bound interfaces so network administrators can write their own network management applications to more effectively run their networks.

\subsection{SDN Datacenter Network}
\begin{figure}[H]
\centering
\includegraphics[scale=0.3]{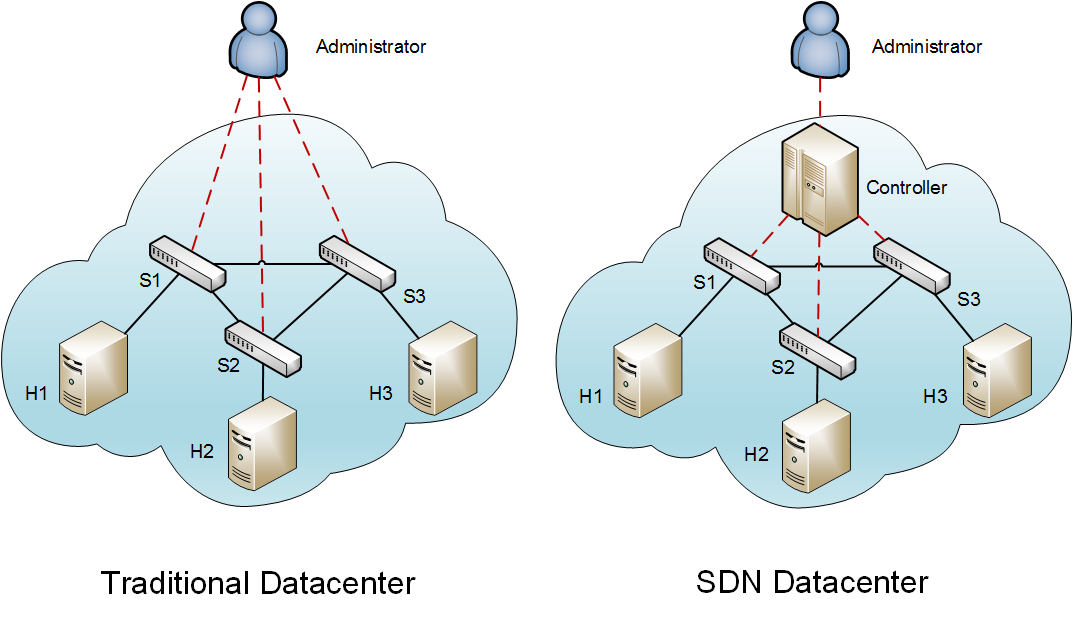}
\caption{SDN/OpenFlow Structure}
\label{fig:sdn-datacenter}
\end{figure}

The decoupled nature of SDN introduces programmability, automaticity and flexibility to the networking management, making SDN a popular solution for large-scale datacenter networks. 
Figure \ref{fig:sdn-datacenter} is a network structure comparison showing the difference between traditional datacenter network and SDN-based datacenter network.

In traditional datacenter network, the administrator needs to configure switches separately. Switches from different vendors with different managing tools become a major obstacle in network management. What's worse, each switch only handles a fragment of the whole network, the lack of global network topology makes it impossible to optimize network traffic dynamically and globally.

Compared with traditional network, SDN reveals great potential in datacenter networks because of its attractive advantages. The SDN controller stores global network topology so that the network can be efficiently optimized. The unified network APIs provided by controller make it easy to perform network management. There have been several successful commercial deployments of SDN in datacenter networks. B4~\cite{B4} network implemented by Google utilizes SDN to schedule network traffic between Google's global datacenters and achieves the maximum link usage of almost $100$ percent; Microsoft deploys SWAN~\cite{SWAN} to dynamically re-configure routing paths of network traffic to optimize inter-datacenter network utilization. Evaluation shows that SWAN can carry 60\% more traffic than traditional network.

\subsection{Information Leakage in Datacenter Network}

In modern datacenter networks like Microsoft Azure and Amazon EC2, customer VMs are usually multiplexed across shared physical infrastructures. Besides achieving a high utilization of hardware and software resources, this approach also introduces new vulnerabilities. Previously published researches have shown that transparently shared physical infrastructures can lead to potential cross-VM information leakage.

At first glance information leakage might seem innocuous, but in fact it is quite useful for clever attackers and will bring security issues in many aspects. The leakage of cache miss information is used in extraction and inference of RSA~\cite{RSA} and AES~\cite{AES} secret keys. The leakage of inter-keystroke time information can be used to perform recovery of the password in keystroke timing attack. The leakage of network topology might provide weapons to attackers because some attacks are only possible when attacker's VM is executed on the same physical server with victim's VM~\cite{Information Leakage}. The leakage of network performance information might make it possible for commercial spies to estimate the number of visitors to a co-resident server belonging to competitors and further infer the operation situation of their company. Thus information leakage in datacenter networks is drawing more and more attentions in network security and privacy researches. 

\subsection{Flow Table Capacity}
Flow table is a hardware structure in OpenFlow switch, it stores hundreds to thousands of routing rules called flow entries. These flow entries are generated and assigned by the controller. Every time a network packet arrives in the switch, the switch will look up its flow table to find corresponding flow entries. If there exists a corresponding flow entry, the switch will forward the network packet according to the actions associated with that flow entry. If there is no flow entry matching this network packet, the switch will send the packet to the controller through the secure channel, then controller will calculate and generate a new flow entry and assign it to the switch.

Previous works typically assume that the flow table of each switch can hold an infinite number of flow entries, which makes the controller easy to design. In practice, however, this assumption does not hold, and the switch flow table capacity can become a significant bottleneck to scaling SDN networks. SDN/OpenFlow switch flow tables often cannot scale beyond a few hundred entries, because they typically include wildcards, and therefore are implemented using either complex and slow data structures, or expensive and extremely power-hungry ternary content-addressable memories (TCAM). Typical SDN/OpenFlow switches have rather limited flow table capacities from 750 to 3000 flow entries while handling about 100,000 concurrent flows in data centers. The flow table capacity bottleneck leads to potential flow table overflow, which is unacceptable.

Combine the flow table capacity issue of SDN switches and the resource sharing phenomenon in SDN-based datacenter networks, we discover the possibility of performing inference attacks targeting at SDN vulnerabilities. A formalized problem statement will be given in next section.
\section{Problem Statement}\label{chap:problem-statement}
Considering the wide use of SDN/OpenFlow in data center networks, we assume the inference attack scene to be in a SDN-based multi-tenant datacenter network like Amazon EC2 or Microsoft Azure. In a SDN-based datacenter network, different tenants connected to the same switch will share the flow table space. The flow table capacity's importance as key network parameter and the possibility of inference attack hidden behind the flow table sharing phenomenon make it natural for us to take flow table capacity as our primary inference target. Besides inferring intrinsic and static property of the switch (flow table capacity), further inference should be performed on the flow table usage condition of other tenants in the same datacenter, which reflects the real-time dynamic resource consuming situation in datacenter networks. So we choose flow table usage as our secondary inference target. But inferring flow table capacity and flow table usage is not that easy.

As cloud computing infrastructures, data center networks are typically well-managed and equipped with advanced firewalls and intrusion detection systems. In order to avoid triggering the IDS, we must behave like an ordinary tenant, which means we cannot gather sensitive information or directly hack into the controller. What's worse, with the constraints of passiveness and concealment, the available parameters for our inference attack are further limited.

After analyzing current structure and implementations of SDN/OpenFlow, its decoupled nature gives us inspiration: the interactions between control plane and data plane will lead to network performance decrease, which can be measured through performance parameters like round trip time (RTT).

\begin{figure}[H]
\centering
\includegraphics[scale=0.58]{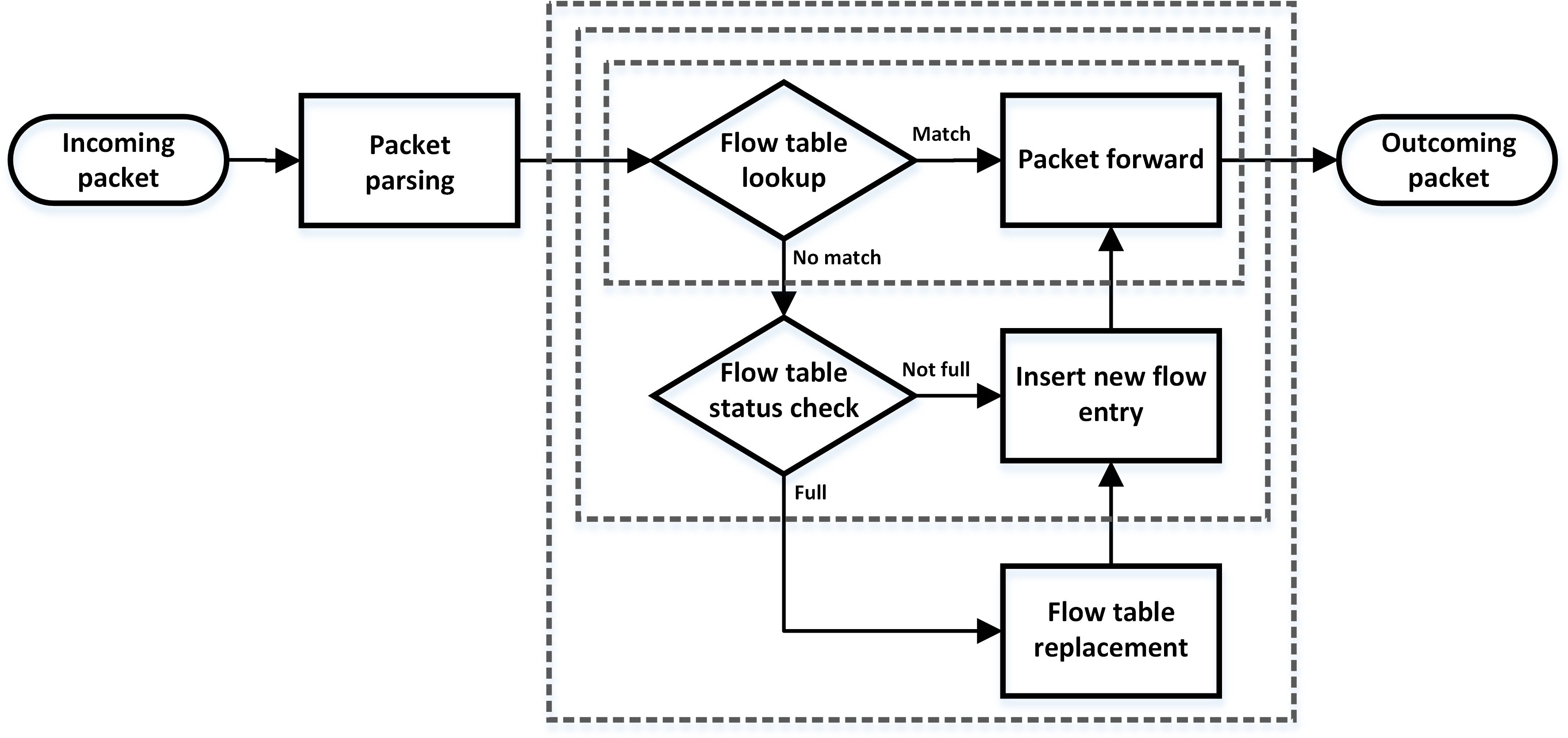}
\caption{OpenFlow Packet Processing Flowchart}
\label{fig:packet-processing-flowchart-of-openflow}
\end{figure}

Fig \ref{fig:packet-processing-flowchart-of-openflow} gives an overall flowchart of packet processing in an OpenFlow switch. The three rectangular regions surrounded by dotted line stand for three possible packet processing branches respectively. When the switch encounters an incoming packet, it will parse it and send the parsed packet into the subsequent processing pipeline.

Then as the first step of the pipeline, the switch will lookup its flow table to search flow entries matching the packet. When there is a match, the switch will directly forward the packet according to actions associated with the corresponding flow entry. This branch is illustrated in the innermost rectangle of fig \ref{fig:packet-processing-flowchart-of-openflow}.

When there is no corresponding flow entry in the flow table, extra steps will be introduced into the procedure. Additional interactions between the switch and the controller will happen to acquire corresponding routing rules, including packet information transferring, routing rule calculation and flow entry deployment. The middle rectangle of fig \ref{fig:packet-processing-flowchart-of-openflow} illustrates this process.

Before the switch inserts the newly generated flow entry, it has to check the flow table status to make sure that there is enough space in the flow table. When the flow table is full, the controller has to perform flow table replacement operations to make room for the upcoming flow entry. These operations include deciding which old flow entry to delete according to certain flow table replacement algorithm and flow entry deletion. The outermost rectangle in fig \ref{fig:packet-processing-flowchart-of-openflow} stands for this branch.

That is exactly where the vulnerability lies. In traditional networks, the switches and routers are autonomous, which means they can maintain their routing tables locally without interacting with an external device. But due to the decoupled nature of SDN/OpenFlow, maintaining switch flow tables needs frequent interactions between switches and controllers, making it possible for an attacker to leverage the perceived performance change to deduce the internal state of the SDN network.

As shown in fig \ref{fig:packet-processing-flowchart-of-openflow}, the rectangular regions surrounded by dotted line correspond to different possible packet processing branches. The larger a rectangle is, the longer the processing time of that branch will be because of the extra steps that rectangle contains. When there is a match in the flow table, the processing time will be the shortest; when there is no match in the flow table and the flow table is not full, the processing time will be longer because of addition routing calculation and flow entry deployment; when there is no match in the flow  table and the flow table is full, the processing time will be the longest because a flow table replacement operation has to be performed. So as a network parameter directly influenced by the processing time, the RTT of a packet can serve as an indicator of flow table state and flow entry state.

\begin{figure}[H]
\centering
\includegraphics[scale=0.45]{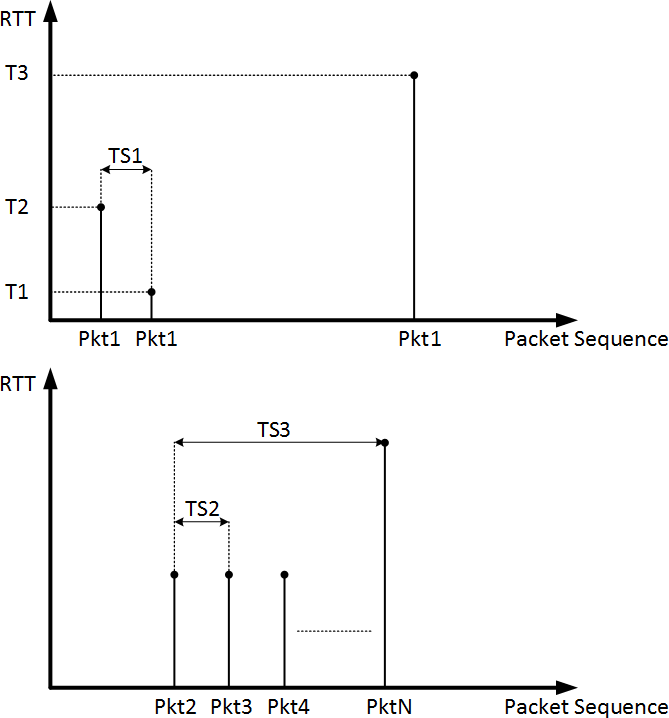}
\caption{RTT Measurement of Different Flow Table State}
\label{fig:rtt-measurement-of-different-flow-table-state}
\end{figure}

The process of deciding RTT thresholds for flow table state detection is shown in figure \ref{fig:rtt-measurement-of-different-flow-table-state}. 

The two sub-figures in figure \ref{fig:rtt-measurement-of-different-flow-table-state} represent two cooperating threads, the x-axis represents the packet sequence and the y-axis represents the recorded RTT of every packet. Firstly, in the upper thread, we generate a packet with a specific \{src\_ip, dst\_ip, src\_mac, dst\_mac\} combination, calling it $Pkt_1$. Send $Pkt_1$ to the target OpenFlow switch and record the corresponding RTT as $T_2$. Currently there is no corresponding flow entry in the OpenFlow switch because $Pkt_1$ is a new packet. After a time span $TS_1$, send $Pkt_1$ to the target OpenFlow switch again and record the corresponding RTT as $T_1$. If $TS_1$ is chosen properly, the newly installed flow entry matching $Pkt_1$ should still exist in the OpenFlow switch. Next, in the lower thread, we continuously generate packets $Pkt_2$, $Pkt_3$, $\cdots$ $Pkt_N$, each with a different combination of \{src\_ip, dst\_ip, src\_mac, dst\_mac\}, and send these packets to the target OpenFlow switch with the time span of $TS_2$. Because there are no flow entries matching there packets in the OpenFlow switch, the recorded RTTs will be approximately the same as $T_2$. Keep generating and sending packets until we observe a sudden increase of the RTT, which indicates that the flow table is full. Then in the upper thread we send $Pkt_1$ again immediately and record the RTT as $T_3$. To achieve higher precision, we can repeat the process and use average values of $T_1$, $T_2$ and $T_3$ as final results.

From the process above we can see $T_1$, $T_2$, $T_3$ will serve as thresholds for flow table state detection: when the measured RTT is around $T_1$, we can infer that there is corresponding flow entry in the flow table; when the measured RTT is around $T_2$, we can infer that there is no corresponding flow entry in the flow table and the flow table is not full; when the measured RTT is around $T_3$, we can infer that there is no corresponding flow entry in the flow table and the flow table is full. The measured RTT thresholds corresponding to different flow table states are shown in table \ref{tab:rtt-comparison}.

\begin{table}[!ht]
\centering
\caption{RTT Comparison}
\label{tab:rtt-comparison}
\begin{tabular}{c c c}
\hline
Flow Table State & Flow Entry State & RTT\\
\hline
NotFull & Exist & $0.2-0.3ms$\\
NotFull & NotExist & $3-5ms$\\
\hline
Full & Exist & $0.2-0.3ms$\\
Full & NotExist & $8-10ms$\\
\hline
\end{tabular}
\end{table}

Having got a method to detect flow table state and flow entry state using the bootstrap process above, our inference model will follow a "probe--observe--infer" pattern. We model the SDN/OpenFlow network as a black box and observe its response (RTT) to different input (network packets), then we use the response to estimate the flow table state and flow entry state and perform further inference. The whole process comes in three steps.

Firstly, we send probing packets into the network to trigger the interaction. As there is still no mature routing aggregation algorithm or hierarchical routing rule solution, current SDN/OpenFlow switches typically use exact-match rules. That means if we send $n$ packets with different faked meta information like src\_ip and dst\_ip, there will be $n$ newly generated flow entries inserted into the flow table. If we send excessive probing packets in a short period of time, the flow table will overflow and then the interaction process will be triggered. Secondly, we measure RTTs of the responded packets and infer the flow table state and flow entry state. Thirdly, we use observed flow table states and flow table states as controlling signals in our inference algorithm and perform flow table capacity inference.

But there is still one problem: how will the flow table deal with the flow entries when the flow table is full. In other words, we need to know the flow table replacement algorithm. The algorithm decides the internal state transition of a SDN network thus it's an essential part of our inference model. Though flow table replacement algorithms of commercial SDN/OpenFlow switches are stored in their firmwares, making it impossible for researchers to read, we can still get illuminations of what the algorithms will be like by analyzing the functionalities of flow tables.

Flow table in a SDN/OpenFlow switch stores the local router tables assigned by the controller. Having to achieve a hit rate as high as possible in a rather limited space, flow table serves like a "cache" in operating systems and web proxy servers. So we have reason to believe that the flow table replacement algorithm will be some of the most popular cache replacement algorithms or their variations. In this paper we choose FIFO\cite{FIFO} and LRU\cite{LRU} because they are common and popular.

So far we have finished our inference model targeting at flow table capacity and there is still another target: flow table usage inference. Though flow table usage inference seems impossible because network traffic is isolated among different tenants, the information leakage attack mentioned in section \ref{chap:background} and our previous analysis provide us with possible breakthrough points: we can infer the flow table capacity and we can record generated flow entries during a time period, the difference of these two values will be the flow table usage during that time period.

Last but not least, we have to discuss the feasibility of our inference attack as well as some key parameters. In order to facilitate the description, we will first introduce the flow entry deletion mechanism of OpenFlow. 

Flow entry deletion mechanism of OpenFlow includes two main parts: active flow deletion invoked by controller and passive flow deletion invoked by timeout. According to OpenFlow specification, each flow entry has two timeout values --- hard\_timeout and idle\_timeout. Hard\_timeout decides how long a flow entry will live after it has been inserted while idle\_timeout means the longest time of no packet matching before a flow entry is deleted. The feasibility of our inference attack will be associated with timeout values and the feasibility analysis consists of two parts: RTT bootstrap feasibility and probing feasibility.

\subsection{RTT Bootstrap Feasibility}
In RTT bootstrap process illustrated in figure \ref{fig:rtt-measurement-of-different-flow-table-state}, there are three key parameters: $TS_1$, $TS_2$ and $TS_3$. Both $TS_1$ and $TS_3$ should not exceed the minimum of hard\_timeout and idle\_timeout because we must prevent the flow entry matching $Pkt_1$ from being deleted in the whole process, which depends on $TS_1$ and $TS_3$ respectively. In order to shorten $TS_3$, $TS_2$ should not be too long because its the time span between every packet and $TS_3$ is made up of $(n-1)$ numbers of $TS_2$. In conclusion, the feasibility constraints for RTT bootstrap are as follows:

\begin{equation}
(n-1)TS_2 = TS_3
\end{equation}

\begin{equation}
TS_1 \leq min\{hard\_timeout, idle\_timeout\}
\end{equation}

\begin{equation}
TS_3 \leq min\{hard\_timeout, idle\_timeout\}
\end{equation}

\subsection{Probing Feasibility}
The key part of our inference model is triggering interactions between switches and controllers by sending probing packets in a short period of time, so the feasibility of our inference attack depends on the feasibility of triggering flow table overflow to a large extent. If we use $V_{gen}$ to represent our packet generating speed and use $V_{del}$ to represent the packet deletion speed and $C$, $T$ to represent the flow table capacity and the probing time, the feasibility formulation will be:

\begin{equation}
V_{gen} \times T - V_{del} \times T \geq C
\end{equation}

Or in another form:

\begin{equation}
V_{gen} \geq V_{del} + C / T
\end{equation}

As can be seen from the formulation, the minimum packet generating speed required are associated with the flow entry deletion speed and flow table capacity. If we set $T$ to be shorter than the minimum of hard\_timeout and idle\_timeout, which means our inference can be completed in a timeout circle, the flow entries deleted during our inference due to timeout will be ignorable. The feasibility constraint will be:

\begin{equation}
V_{gen} \geq C / min\{hard\_timeout, idle\_timeout\}
\end{equation}

From the feasibility constraints above, we can conclude that the feasibility of our inference attack depends on the timeout measurement. It is essential for us to measure hard\_timeout and idle\_timeout not only for inference time limit but also for packet generating speed adjustment.

\begin{figure}[H]
\centering
\includegraphics[scale=0.45]{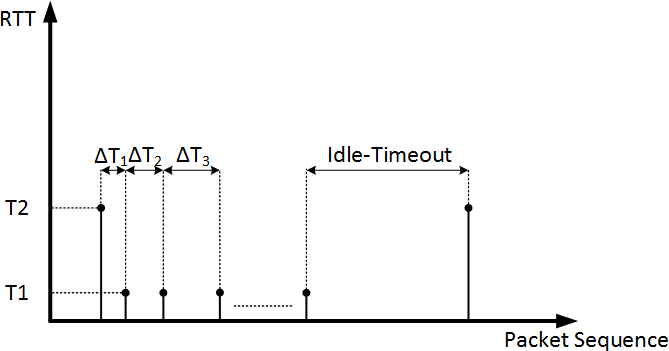}
\caption{Idle\_timeout Measurement}
\label{fig:idle-timeout-measurement}
\end{figure}

Idle\_timeout can be measured by sending a train of packets with gradually increasing packet intervals. The process is shown in Figure \ref{fig:idle-timeout-measurement}.

We choose an initial value of packet interval $\Delta T_1$, for example $0.1$s. Then we gradually increase the packet interval to $\Delta T_2$, $\Delta T_3$ and so on using "binary search" or other algorithms. Keep sending these packets with increasing intervals until we observe a significantly high RTT value. The corresponding packet interval at that time will be the idle\_timeout.

\begin{figure}[H]
\centering
\includegraphics[scale=0.45]{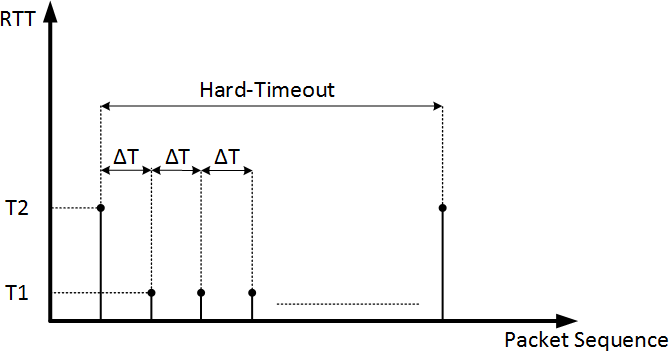}
\caption{Hard\_timeout Measurement}
\label{fig:hard-timeout-measurement}
\end{figure}

Hard\_timeout can be measured by sending a train of packets with constant packet intervals far smaller that the previously measured idle\_timeout. Keep sending packets with a packet interval of $\Delta T$, for example $0.1$s. When observing a significantly high RTT value, the corresponding packet interval will be the hard\_timeout. The process is shown in Figure \ref{fig:hard-timeout-measurement}.

In this section we set our inference targets to be flow table capacity and flow table usage, we also discussed the inference model and its feasibility. In section \ref{chap:fifo-inference-algorithm} and \ref{chap:lru-inference-algorithm} we'll give detailed descriptions of inference algorithms for FIFO and LRU respectively.
\section{FIFO Inference Algorithm}\label{chap:fifo-inference-algorithm}

\begin{figure}[H]
\centering
\includegraphics[scale=0.3]{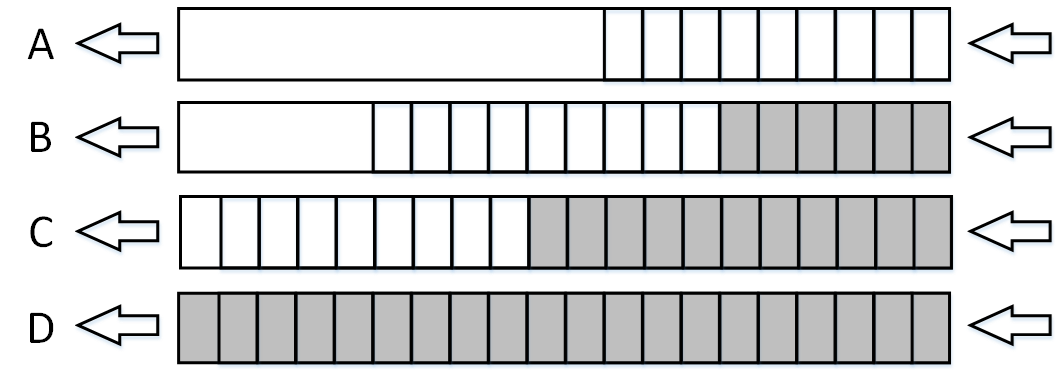}
\caption{FIFO Inference Principle}
\label{fig:fifo-step}
\end{figure}

As mentioned in section \ref{chap:problem-statement}, the inference process of FIFO algorithm will be as follows: we generate and send a huge amount of probing packets each with a different combination of {src\_ip, dst\_ip, src\_mac, dst\_mac}, the newly inserted flow entries matching the generated packets will "push" the other users' flow entries out of the flow table. We can detect if the flow table is full and the existence of our flow entries. Combined with the number of inserted flow entries we recorded, we can infer the flow table capacity and flow table usage. The process of flow table state transformation is shown in figure \ref{fig:fifo-step}.

We use $F_{our}$ to represent the number of our inserted flow entries and use $F_{other}$ to represent the number of flow entries from other users in the flow table. Both $F_{our}$ and $F_{other}$ are functions of time. We use $T_A$, $T_B$, $T_C$ and $T_D$ to represent four time points corresponding to four sub figures respectively and use $C$ to respresent the flow table capacity.

Figure \ref{fig:fifo-step} (A) shows the flow table and the flow entries it contains just before the experiment starts. The rectangle items represent the flow entries from other users sharing the OpenFlow switch. The current number of other users' flow entries can be expressed as $F_{other}(T_A)$.

Figure \ref{fig:fifo-step} (B) illustrates the time when we start to send generated packets, inserting new flow entries into the flow table. The grey rectangles represent the flow entries inserted by us. As we can see, our flow entries keep pushing other users' flow entries to the front of the FIFO queue. During the experiment, we should keep a record of the generated packets, including their attributes and serial numbers.

Figure \ref{fig:fifo-step} (C) shows the time when we detect the flow table is full. At this point of time, flow entries from us and other users add up to fill the whole flow table precisely. We have:

\begin{equation}
F_{our}(T_C) + F_{other}(T_C) = C
\end{equation}

Figure \ref{fig:fifo-step} (D) shows the time when we detect that one of our inserted flow entries has been deleted. That means the flow table is now full of our flow entries, without any flow entries from other users. We have:
\begin{equation}
F_{our}(T_D) = C
\end{equation}

Combine the two equations above, we have:

\begin{equation}
\begin{aligned}
&F_{other}(T_A) = F_{other}(T_C) \\
&= C - F_{our}(T_C) = F_{our}(T_D) - F_{our}(T_C)
\end{aligned}
\end{equation}

According to the analysis above, we describe the inference process for FIFO algorithm as shown below.

\begin{algorithm}[H]
\caption{FIFO Inference Algorithm}
\begin{algorithmic}[1]
\Require\\
Packet-Sending Function: $SendPacket\left(\right)$;\\
List of IP: $IP$;
\Ensure\\
The flow table capacity: $F_{capacity}$;\\
The number of other users' flow entries: $F_{other}$;
\Statex
\State $F_{capacity} \gets 0$
\State $F_{other} \gets 0$
\State $N \gets 0$
\State $N_1 \gets 0$
\State $N_2 \gets 0$
\While {$N<length(IP)$}
\State $ip \gets IP[N]$
\State $\Call{SendPacket}{ip}$
\State $N \gets N+1$
\If {Flow table is full}
\State $N_1 \gets N$
\State continue
\EndIf
\If {One of our flow entries is deleted}
\State $N_2 \gets N$
\State break
\EndIf
\EndWhile
\State $F_{capacity} \gets N_2$
\State $F_{other} \gets N_2-N_1$
\State \Return {$F_{capacity}$, $F_{other}$}
\end{algorithmic}
\end{algorithm}

The main error of the inference comes from the flow entries inserted by other users when our insertion is in progress. We assume that our flow entry insertion speed is fast enough so that during the period of experiment, the newly inserted flow entries are all from us. But that is not always the truth. Ignoring the possible flow entries inserted by other users will make our inference result smaller than the actual value.

Considering the flow entries inserted by other users, the actual equations are listed below.

When we detect the flow table is full, if we use $E(A, B)$ to represent the number of just inserted flow entries from other users from time point $A$ to time point $B$, the equation becomes:
\begin{equation}
F_{our}(T_C) + F_{other}(T_C) + E(T_A, T_C) = C
\end{equation}

And when we detect one of our inserted flow entries is deleted, the equation becomes:
\begin{equation}
F_{our}(T_D) + E(T_A, T_C) + E(T_C, T_D) = C
\end{equation}

Combine the two equations above, we have:
\begin{equation}
F_{other}(T_C) = F_{our}(T_D) - F_{our}(T_C) + E(T_C, T_D)
\end{equation}

So the actual equation considering flow entry insertions during inference should be:

\begin{equation}
\begin{aligned}
C &= F_{our}(T_D) + E(T_A, T_C) + E(T_C, T_D) \\
F_{other}(T_C) &= F_{our}(T_D) - F_{our}(T_C) + E(T_C, T_D)
\end{aligned}
\end{equation}

Compared with our former equation ignoring flow entry insertions:

\begin{equation}
\begin{aligned}
C &= F_{our}(T_D) \\
F_{other}(T_C) &= F_{our}(T_D) - F_{our}(T_C)
\end{aligned}
\end{equation}

We can see that the inferred flow table usage $F_{other}$ and the inferred flow table capacity $F_{capacity}$ will both be smaller than the actual value.
\section{LRU Inference Algorithm}\label{chap:lru-inference-algorithm}
The experiment principle of LRU algorithm has something in common with that of FIFO algorithm, because under these two circumstances we can both keep our flow entries stay in the back of the cache queue using certain operations.However, there are still differences lies in the flow entry maintaining process.

The nature of FIFO algorithm ensures that the position of the flow entries only depends on the time they are inserted. The earlier inserted flow entries are sure to be nearer to the front of the cache queue compared with the later inserted flow entries. But in LRU algorithm, the positions of the flow entries depend not only on the time they are inserted, but also on the last time they are accessed. In order to keep our flow entries stay in the back of the cache queue, we need to continuously access the previously inserted flow entries.

During the maintain process, every time we insert a new flow entry, we need to access all previously inserted flow entries for one time to "lift" them to the back of the cache queue. The access history may be like $\{P_1\},\ \{P_1,\ P_2\},\ \{P_1,\ P_2,\ P_3\},\ \{P_1,\ P_2,\ P_3,\ P_4\}, \cdots$, we call it a "rolling" maintaining process. The maintaining algorithm is shown in algorithm 2.

\begin{algorithm}[H]
\caption{Rolling Maintaining Algorithm}
\begin{algorithmic}[1]
\Require\\
Packet-Sending Function: $SendPacket()$;\\
List of Inserted IP: $IP_{inserted}$;
\Statex
\Function {RollingPacketSender}{$IP_{inserted}$}
\State $i \gets 1$
\While {$i < length(IP_{inserted})$}
\For {$j \gets 0;j < i;j++$}
\State $ip \gets IP_{inserted}[j]$
\State \Call{SendPacket}{$ip$}
\EndFor
\State $i \gets i+1$
\EndWhile
\EndFunction
\end{algorithmic}
\end{algorithm}

According to the analysis above, we describe the inference process for LRU algorithm as shown below.

\begin{algorithm}[H]
\caption{LRU Inference Algorithm}
\begin{algorithmic}[1]
\Require\\
Packet-Sending Function: $SendPacket\left(\right)$;\\
List of IP: $IP$;
\Ensure\\
The flow table capacity: $F_{capacity}$;\\
The number of other users' flow entries: $F_{other}$;
\Statex
\State $F_{capacity} \gets 0$
\State $F_{other} \gets 0$
\State $N \gets 0$
\State $N_1 \gets 0$
\State $N_2 \gets 0$
\State $IP_{inserted} \gets [\ ]$
\While {$N<length(IP)$}
\State $ip \gets IP[N]$
\State $IP_{inserted} \gets IP_{inserted} + ip$
\State \Call{RollingPacketSender}{$IP_{inserted}$}
\State $N \gets N+1$
\If {Flow table is full}
\State $N_1 \gets N$
\State continue
\EndIf
\If {One of our flow entries is deleted}
\State $N_2 \gets N$
\State break
\EndIf
\EndWhile
\State $F_{capacity} \gets N_2$
\State $F_{other} \gets N_2-N_1$
\State \Return {$F_{capacity}$, $F_{other}$}
\end{algorithmic}
\end{algorithm}

The feasibility and error analysis of LRU algorithm is similar with that of FIFO algorithm. The inferred flow table usage $F_{other}$ and the inferred flow table capacity $F_{capacity}$ will both be smaller than the actual value because of ignoring the flow entries inserted by other users during the experiment.
\section{Evaluation}\label{chap:evaluation}

\subsection{Implementation}
The emulation environment of our experiment consists of three parts: a network prototyping system used to emulate host and switch, a network controller, and our inference attack toolkit.

We choose Mininet~\cite{Mininet} as the network prototyping system because it encapsulates host and switch emulation and thus easy to use. Our emulated network prototype for evaluation uses a star topology, consisting of $20$ hosts connected to a single OpenFlow switch. We build FIFO and LRU controller applications using Python on the basis of POX~\cite{POX} OpenFlow controller. As for the inference attack toolkit, we use libnet~\cite{libnet} to generate probing packets, and libpcap~\cite{libpcap} to capture replied packets. Experiments are conducted on a computer with Intel i5-2400 3.1 GHz (4 cores) processor and 8GB RAM.

\subsection{RTT Measurement}
As we have mentioned in section \ref{chap:problem-statement}, the difference between traditional network and SDN/OpenFlow network in handling previously unseen packets gives us a possible indicator of the flow table state and the flow entry living state -- RTT. When there isn't corresponding flow entry existing in the flow table, the RTT of a packet will significantly increase due to the interactions between controller and switch in order to acquire new flow entries. That is the case when there is still space in the flow table. Once the flow table is full, the RTT of a packet will further increase as a result of extra flow table replacement operations. To prove the effectiveness of using RTT as the flow table state and flow entry state indicator, we measured packet RTTs corresponding to different flow table state and flow entry state.

\begin{figure}[H]
\centering
\includegraphics[scale=0.3]{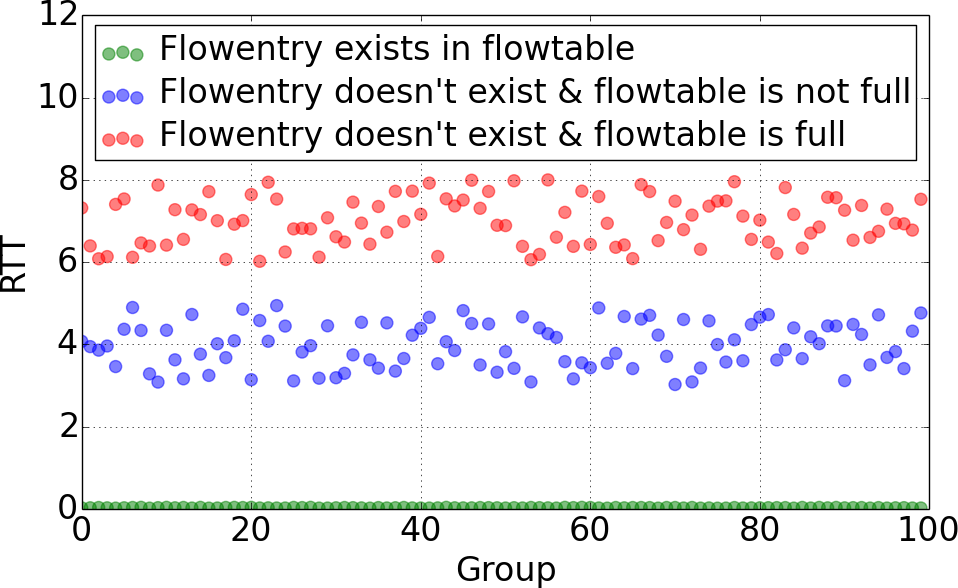}
\caption{RTT Measurement}
\label{fig:rtt-measurement}
\end{figure}

Figure \ref{fig:rtt-measurement} gives the RTT measurement result showing the difference. The points with different colors represent the total $300$ times of RTT measurements we have conducted, $100$ times of measurement for each combination of flow table state and flow entry state. The green points stand for RTTs when flow entry exists in flow table. The blue points and red points both stand for RTTs when flow entry doesn't exist in flow table, the only difference is the flow table is not full when measuring the blue points.

As can be seen from the figure, when flow entry exists in flow table, the packet RTTs are highly concentrated in the range of $0.2\sim0.3$ ms; when flow entry doesn't exist in flow table and flow table is not full, the packet RTTs will increase to about $3\sim5$ ms; when flow entry doesn't exist in flow table and flow table is full, the packet RTTs will be the highest, ranging from $6$ms to $8$ms. These three groups of RTTs all distribute intensively in a small range without overlapping other groups, showing the excellent discrimination of using RTT as a flow table state and flow entry state indicator.

\begin{figure}[H]
\centering
\includegraphics[scale=0.3]{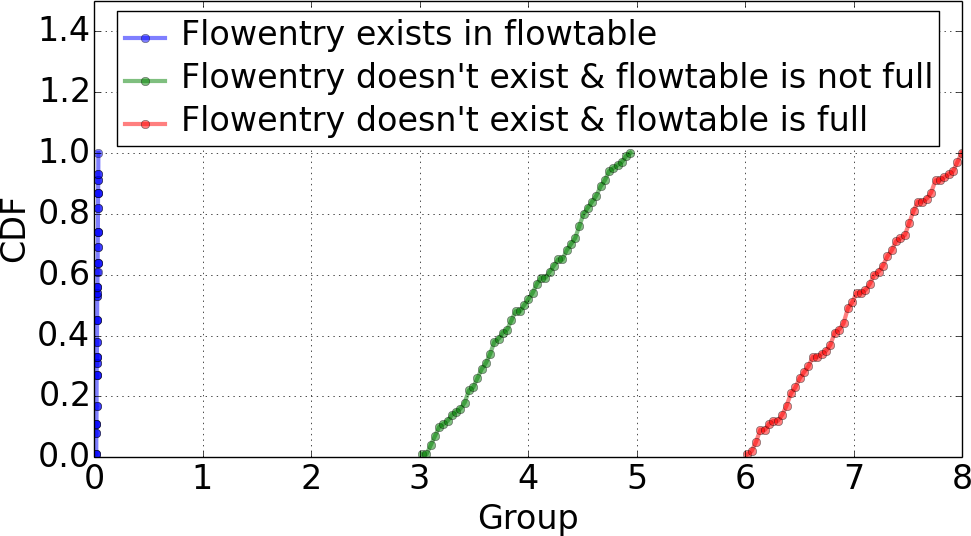}
\caption{RTT Measurement Distribution}
\label{fig:rtt-measurement-distribution}
\end{figure}

To better illustrate the distribution of measured RTTs, we plot their CDF curves in figure \ref{fig:rtt-measurement-distribution}. Apparently RTT can be used to deduce the internal state of the SDN network effectively.

\subsection{Timeout}

\subsubsection{Default Timeout Values}

According to our previous analysis, the feasibility of our inference attack depends on whether we can generate enough flow entries to fulfill the flow table within a single timeout cycle. That means we must have the ability to generate as many flow entries as the flow entry can hold during a timeout period. So we analyze several popular open-source controllers and search for their default timeout values in the built-in applications. The result is presented in table \ref{tab:default-timeout-values}. The zero values in the table mean the corresponding timeout will not take effect, or in other words the timeout value is "permanent". As can be seen from the table, most available controllers have timeout values in the range of $5$s to $30$s.

\begin{table}[!ht]
\centering
\caption{Default Timeout Values}
\label{tab:default-timeout-values}
\begin{tabular}{l c c}
\hline
Controller & Hard\_timeout & Idle\_timeout\\
\hline
Ryu & $0$ & $0$\\
Beacon & $0$ & $5s$\\
Floodlight & $0$ & $5s$\\
NOX & $0$ & $5s$\\
POX & $30s$ & $10s$\\
Trema & $0$ & $60s$\\
Maestro & $180s$ & $30s$\\
\hline
\end{tabular}
\end{table}

If we take the flow table capacity of $2000$ flow entries as an example, the minimum packet generating speed required will be $2000/5=400$ packets per second, while libnet can generate tens of thousand packets per second. So the default timeout values ensure the feasibility of our inference attack. 

\subsubsection{Timeout Measurement}
Though default timeout values of mainstream OpenFlow controllers can be read from their source codes, it's still possible for SDN network administrators to manually change the default timeout values. In order to handle non-default timeout values and provide basis for adjusting packet generating speed, it's essential to examine the accuracy of passive timeout measurement.

\begin{figure}[H]
\centering
\includegraphics[scale=0.23]{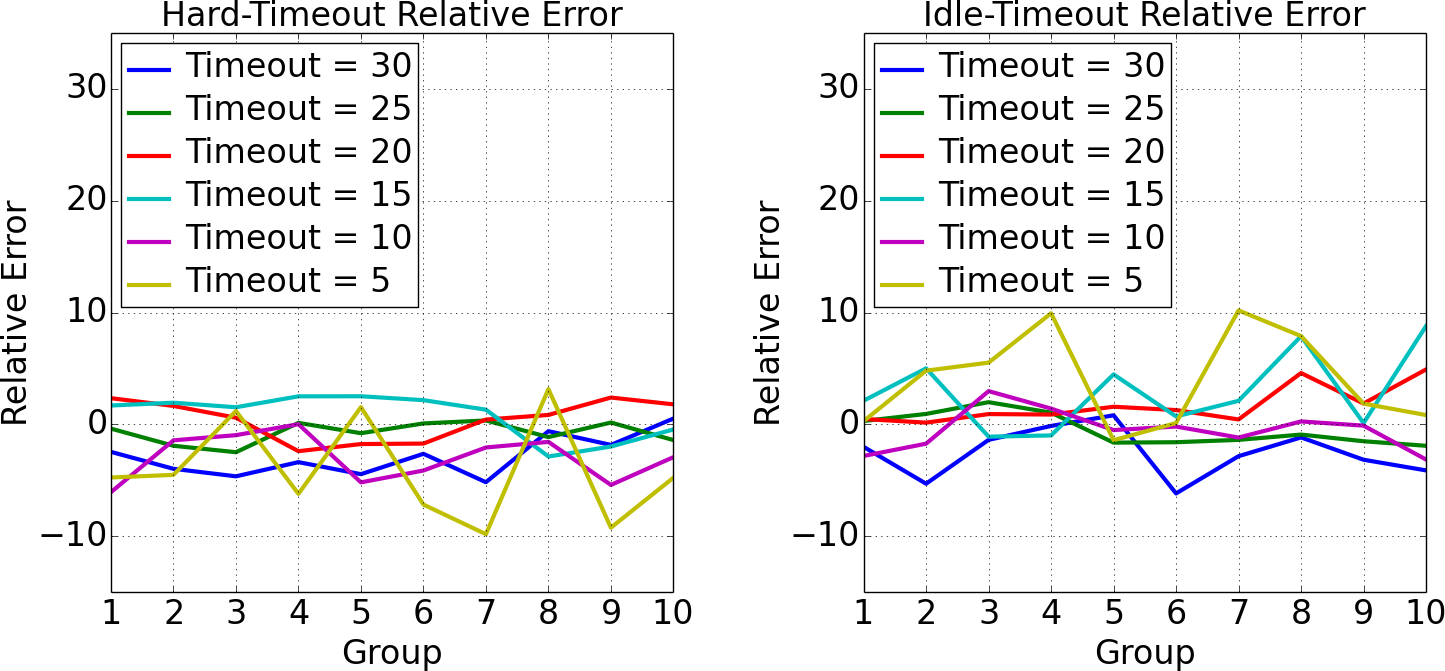}
\caption{Timeout Relative Error}
\label{fig:timeout-relative-error}
\end{figure}

Figure \ref{fig:timeout-relative-error} illustrates relative errors of hard\_timeout and idle\_timeout measurement respectively. We manually modify hard\_timeout and idle\_timeout values of POX OpenFlow controller to $5$s, $10$s, $15$s, $20$s, $25$s and $30$s, then we use timeout measurement algorithm mentioned in section \ref{chap:problem-statement} to measure these timeout values and calculate relative errors.

Every line in the two sub-figures corresponds to $10$ times of repeated measurements conducted under a certain timeout setting from $5$s to $30$s. The margin stays in the range of plus-or-minus $10$ percent, showing the effectiveness and high accuracy of our timeout measurement algorithm.

\subsection{Flow Table Capacity}

\begin{figure}[H]
\centering
\includegraphics[scale=0.25]{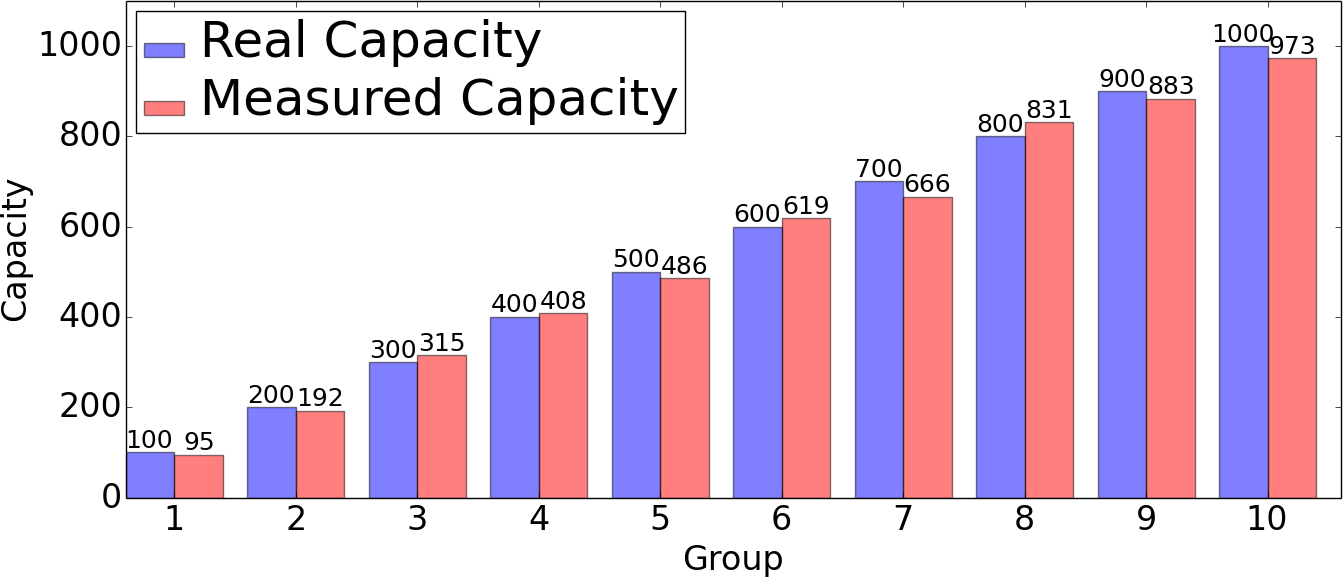}
\caption{FIFO Flow Table Capacity}
\label{fig:fifo-capacity}
\end{figure}

Flow capacity is the primary target of our inference attack. It reflects the hardware specification of an OpenFlow switch. Figure \ref{fig:fifo-capacity} illustrates the flow table capacity measurement result when controller adopts FIFO replacement algorithm. We manually limited the switch flow table capacity to $10$ different values from $100$ flow entries to $1000$ flow entries and used our framework to perform the inference.

The pink bars represent the manually set flow table capacities or \textit{real} capacities. The blue bars represent the measured flow table capacities. For every manually set flow table capacity, we conduct $10$ times of repeated measurements and take their mean value as the final result. From the figure we can see that the measured capacities is quite close to the real capacities, indicating the high accuracy of our inference framework. For example, when the real capacity is $400$ flow entries, our measured capacity is $408$ flow entries with an error of only $8$ flow entries. As the real capacity grows, the packet generating speed required becomes faster, placing higher requirements on packet sending -- receiving synchronization and accurate timing. But our inference algorithm shows unbelievable stability and accuracy: when the real capacity is $1000$ flow entries, our measured capacity is $973$ flow entries with an error of just $27$ flow entries.

\begin{figure}[H]
\centering
\includegraphics[scale=0.25]{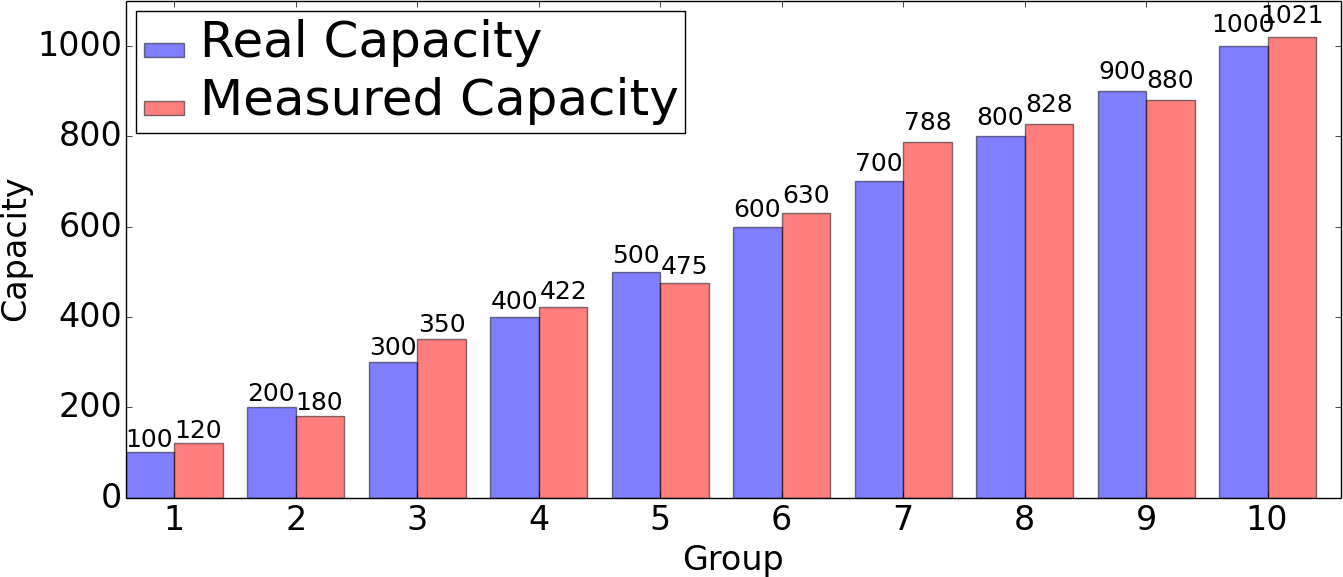}
\caption{LRU Flow Table Capacity}
\label{fig:lru-capacity}
\end{figure}

Like figure \ref{fig:fifo-capacity}, figure \ref{fig:lru-capacity} also illustrates the flow table capacity measurement results, with the only difference of being performed under LRU replacement algorithm instead of FIFO.

According to our previous analysis, the inference principle of LRU replacement algorithm is more complex because of the unavoidable mixed nature of flow entries in the flow table and the rolling maintaining process. But our inference framework still shows high accuracy and reliability. Even when the real flow table capacities are set to be rather large values like $900$ and $1000$, the errors of our measure capacities are just around $20$ flow entries.

\begin{figure}[H]
\centering
\includegraphics[scale=0.23]{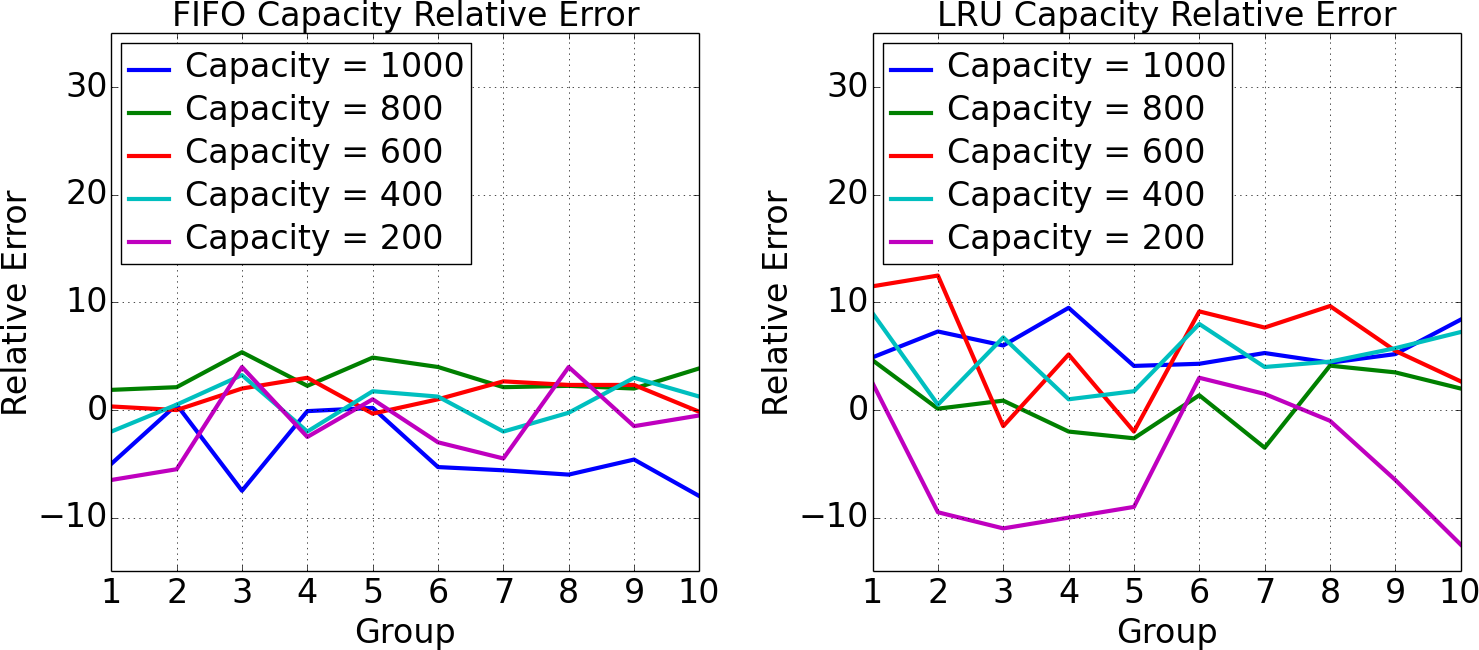}
\caption{Flow Table Capacity Relative Error}
\label{fig:flow-table-capacity-relative-error}
\end{figure}

Only illustrating the mean value of measured flow table capacities may not be enough: the mean value may be the result of error compensations and hide the detailed measurement errors of every separate experiment. So in figure \ref{fig:flow-table-capacity-relative-error} we illustrates the relative error of every single flow table capacity measurement.

We choose $5$ groups of different flow table capacities from $200$ flow entries to $1000$ flow entries and perform $10$ times of measurements under every single flow table capacity value. The left sub-figure stands for relative error of flow table capacity measurements conducted under FIFO replacement algorithm, showing that the margin is no larger than plus-or-minus $10$ percent. The right sub-figure stands for relative error of flow table capacity measurements conducted under LRU replacement algorithm. Due to the more complex inference principle and the rolling maintaining process, the margin becomes larger but still hasn't exceeded $15$ percent even in the worst case.

\subsection{Flow Table Usage}

\begin{figure}[H]
\centering
\includegraphics[scale=0.25]{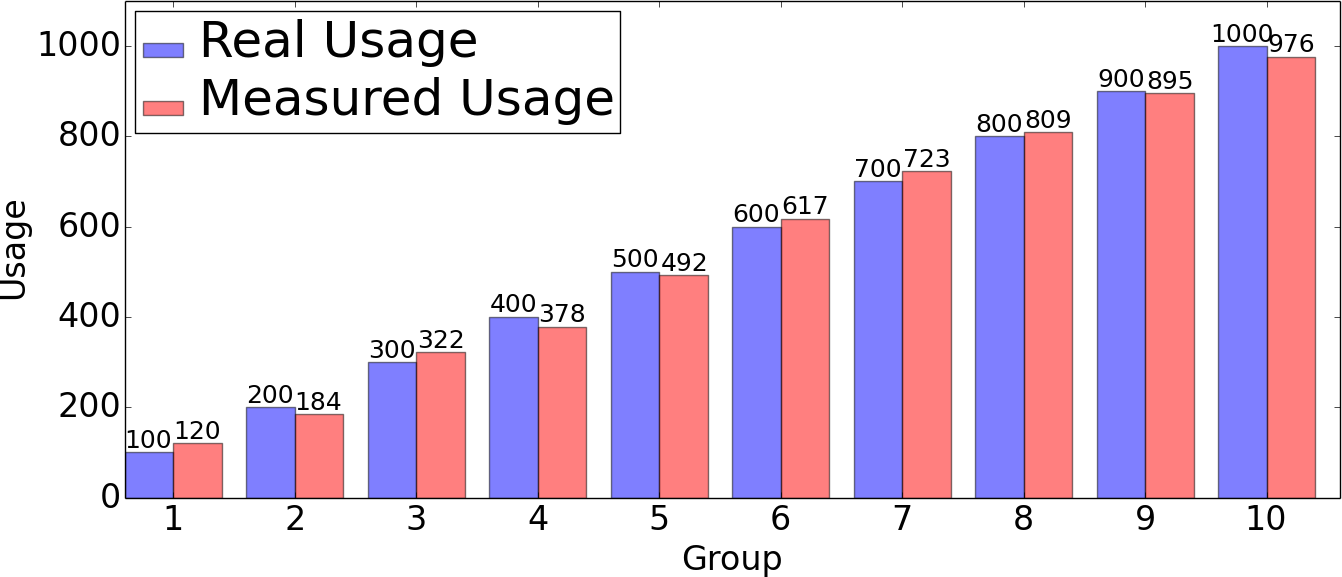}
\caption{FIFO Flow Table Usage}
\label{fig:fifo-usage}
\end{figure}

\begin{figure}[H]
\centering
\includegraphics[scale=0.25]{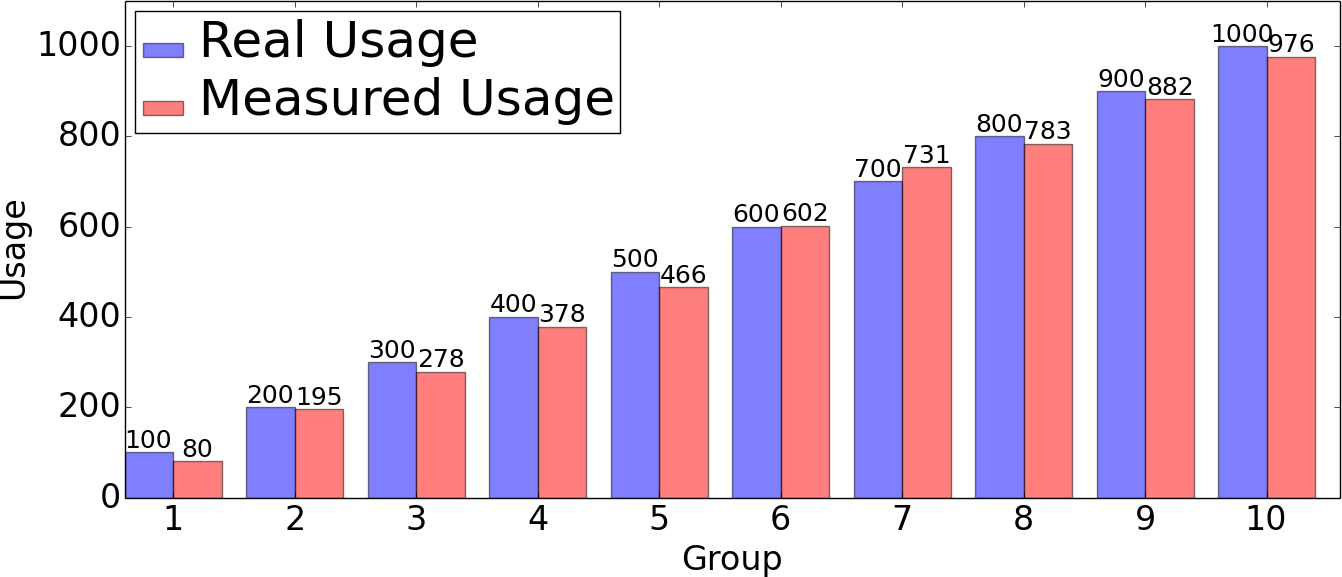}
\caption{LRU Flow Table Usage}
\label{fig:lru-usage}
\end{figure}

In this section we evaluated our framework's efficiency of inferring the number of flow entries from other users sharing the same flow table, or the flow table usage. Flow table usage is our secondary inference target, it reflects the network resource consuming condition of other tenants in the same SDN network. Figure \ref{fig:fifo-usage} and figure \ref{fig:lru-usage} illustrate the flow table usage measurement results conducted under FIFO and LRU replacement algorithm respectively.

Again we manually set $10$ different flow table usage values from $100$ to $1000$ flow entries by manually generating and inserting corresponding number of flow entries into the flow table beforehand. Then we use our inference algorithm to infer the flow table usage and take mean values of every $10$ times of measurements as the final results. The errors of all these measurements show the high accuracy, stability and reliability of our inference algorithm. 

\begin{figure}[H]
\centering
\includegraphics[scale=0.23]{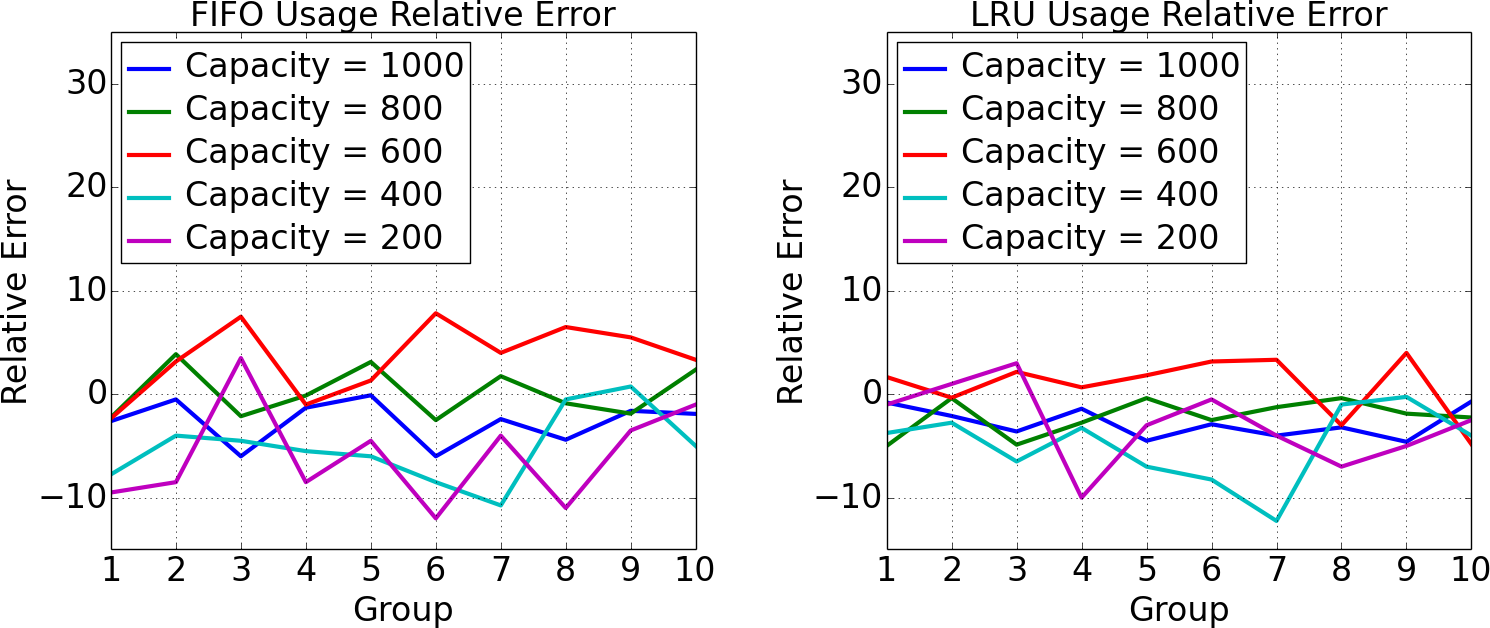}
\caption{Flow Table Usage Relative Error}
\label{fig:flow-table-usage-relative-error}
\end{figure}

The relative errors are shown in figure \ref{fig:flow-table-usage-relative-error}. We emulate $5$ groups of different flow table usage values and conducted $10$ times of flow table usage inference for every single value. For both FIFO and LRU replacement algorithm, the relative errors of flow table usage inference stay in a quite small range. The results prove that our algorithm can infer other tenants' flow table usage condition in high accuracy.

\section{Discussion}\label{chap:discussion}
SDN/OpenFlow has become a competitive solution for next generation network and is being more and more widely used in modern datacenters. But considering its key role as the datacenter fundamental infrastructure, we have to admit that the security issues of SDN/OpenFlow haven't been explored to a large extent. Especially, the flow table capacity of SDN/OpenFlow switch is only considered as a vulnerable part for DDoS and flooding attacks in published researches. But according to our analysis in previous sections, the flow table capacity can lead to potential inference attack if combined with reasonable assumptions and RTT measurements.

Firstly, we found in section \ref{chap:problem-statement} that exact match flow entries as well as the lack of route aggregation would consume a lot of flow table space, making it impossible to process millions of flows per seconding using SDN/OpenFlow. Secondly, we found in section \ref{chap:evaluation} that assigning the decision making job of flow table replacement to the controller would lead to significant network performance decrease, which had to be changed in time. Thirdly, there is currently no mature attack detection mechanism for SDN/OpenFlow network, so it's quite easy for criminals to exploit system vulnerabilities or invoke DDoS attacks.

All these security issues call for improvements to current OpenFlow switch and flow table design. The improvements should at least contain three aspects: (1)new flow table maintaining mechanism, like transferring the flow entry deleting workload from controller to switch. Switch itself can decide which flow entry to delete and then sync with controller. In the newest OpenFlow switch specification 1.4~\cite{OpenFlow Switch Specification 1.4}, this mechanism has been added as an optional feature, but without any mature implementation so far; (2) routing aggregation. Routing aggregation can match a group of flows using one flow entry, which will reduce the flow table consuming significantly compared with exact match; (3) inference attack detection. Administrators can develop inference attack detecting applications and then perform defences like port speed limiting or network address validation.
\section{Related Work}\label{chap:related-work}
The inference attack proposed in this paper is motivated by the limited flow table capacity of SDN/OpenFlow switches. The flow table capacity issue has been presented in many previous works like~\cite{Sezer},~\cite{Kreutz} and~\cite{Scott}. They all point out the limitation of switch flow table memory and potential scalability and security issue. However, these work don't give further analysis on the inference attack and information leakage caused by the limited flow table capacity.

Kloti et al.~\cite{Kloti} presents potentially problematic issues in SDN/OpenFlow including information disclosure through timing analysis. However, this information disclosure requires disclosing existing flows with side channel attack, which is hard to perform in real world. Compared with their approach, our inference attack is self-contained and requires no prior knowledge. 

Gong et al.~\cite{Gong} presents a kind of inference attack using RTT measurement to infer which website the victim is browsing. They recover victims' network traffic patterns based on the queuing side channel happened at the Internet router. However, the scenario of their work is in the public Internet, while our approach focus on SDN/OpenFlow infrastructures in datacenter network. Compared with public Internet and website inference, the inference attack and information leakage in modern data centers is more sensitive and valuable.

Shin et al.~\cite{Shin} demonstrate a novel attack targeting at SDN networks. This attack includes fingerprinting SDN networks and further flooding the data plane flow table by sending specifically crafted fake flow requests in high speed. In the fingerprinting phase, header field change scanning is used to collect the different response time (RTT) for new flow and existing flow. The fingerprinting result is then analyzed to estimate if the target network used SDN technology. The RTT measurement and analysis they used in fingerprinting is similar with our approach. But they just perform DoS attacks to the SDN network, without performing any further information leakage or network parameter inference.

\section{Conclusion}\label{chap:conclusion}
In this paper, we have explored the structure of SDN/\\OpenFlow network and some of the possible security issues it brings. After out detailed analysis of the SDN/\\OpenFlow network, we proposed a novel inference attack model targeting at the SDN/OpenFlow network, which is the first proposed inference attack model of this kind in the SDN/OpenFlow area. This inference attack is introduced by the OpenFlow switch, especially by its limited flow table capacity. The inference attack can be done in a completely passive way, making it hard to detect and defence. We also implemented the inference attack framework and examined the efficiency and accuracy of it using network traffic data from different sources. The simulation results show that the inference attack framework can infer the network parameter(flow table capacity and flow table usage) with an accuracy of up to $80\%$ or higher.

\section{Acknowledgment}
The authors wish to thank the anonymous reviewers for their helpful feedback. The research presented in this paper is supported in part by the National Natural Science Foundation (61221063, U1301254), 863 High Tech Development Plan (2012AA011003) and 111 International Colaboration Program of China.

\bibliographystyle{abbrv}

\begin{thebibliography}{1}

\bibitem{OpenFlow Vulnerability Assessment}
Benton, Kevin, L. Jean Camp, and Chris Small. "Openflow vulnerability assessment." Proceedings of the second ACM SIGCOMM workshop on Hot topics in software defined networking. ACM, 2013.

\bibitem{FortNOX}
Porras, Philip, et al. "A security enforcement kernel for OpenFlow networks." Proceedings of the first workshop on Hot topics in software defined networks. ACM, 2012.

\bibitem{NOX}
Gude, Natasha, et al. "NOX: towards an operating system for networks." ACM SIGCOMM Computer Communication Review 38.3 (2008): 105-110.

\bibitem{VeriFlow}
Khurshid, Ahmed, et al. "Veriflow: Verifying network-wide invariants in real time." ACM SIGCOMM Computer Communication Review 42.4 (2012): 467-472.

\bibitem{PAST}
Stephens, Brent, et al. "PAST: Scalable Ethernet for data centers." Proceedings of the 8th international conference on Emerging networking experiments and technologies. ACM, 2012.

\bibitem{Network traffic characteristics of data centers in the wild}
Benson, Theophilus, Aditya Akella, and David A. Maltz. "Network traffic characteristics of data centers in the wild." Proceedings of the 10th ACM SIGCOMM conference on Internet measurement. ACM, 2010.

\bibitem{Flow caching for high entropy packet fields}
Shelly, Nick, et al. "Flow caching for high entropy packet fields." Proceedings of the third workshop on Hot topics in software defined networking. ACM, 2014.

\bibitem{CAB}
Yan, Bo, et al. "CAB: a reactive wildcard rule caching system for software-defined networks." Proceedings of the third workshop on Hot topics in software defined networking. ACM, 2014.

\bibitem{CacheFlow}
Katta, Naga, Jennifer Rexford, and David Walker. "Infinite cacheflow in software-defined networks." Princeton School of Engineering and Applied Science, Tech. Rep (2013).

\bibitem{OpenFlow Introduction}
McKeown, Nick, et al. "OpenFlow: enabling innovation in campus networks." ACM SIGCOMM Computer Communication Review 38.2 (2008): 69-74.

\bibitem{B4}
Jain, Sushant, et al. "B4: Experience with a globally-deployed software defined WAN." Proceedings of the ACM SIGCOMM 2013 conference on SIGCOMM. ACM, 2013.

\bibitem{SWAN}
Hong, Chi-Yao, et al. "Achieving high utilization with software-driven WAN." Proceedings of the ACM SIGCOMM 2013 conference on SIGCOMM. ACM, 2013.

\bibitem{RSA}
Percival, Colin. "Cache missing for fun and profit." (2005).

\bibitem{AES}
Osvik, Dag Arne, Adi Shamir, and Eran Tromer. "Cache attacks and countermeasures: the case of AES." Topics in Cryptology–CT-RSA 2006. Springer Berlin Heidelberg, 2006. 1-20.

\bibitem{Information Leakage}
Ristenpart, Thomas, et al. "Hey, you, get off of my cloud: exploring information leakage in third-party compute clouds." Proceedings of the 16th ACM conference on Computer and communications security. ACM, 2009.

\bibitem{FIFO}
"FIFO" http://en.wikipedia.org/wiki/FIFO

\bibitem{LRU}
"LRU" http://en.wikipedia.org/wiki/LRU

\bibitem{Mininet}
"Mininet" http://mininet.org/

\bibitem{POX}
"POX Controller" http://www.noxrepo.org/pox/about-pox/

\bibitem{libnet}
"libnet-dev" https://github.com/sam-github/libnet

\bibitem{libpcap}
"libpcap" http://www.tcpdump.org/

\bibitem{OpenFlow Switch Specification 1.4}
"OpenFlow Switch Specification 1.4.0" https://www.opennetworking.org/images/stories\\/downloads/sdn-resources/onf-specifications/openflow/openflow-spec-v1.4.0.pdf

\bibitem{Sezer}
Sezer, Sakir, et al. "Are we ready for SDN? Implementation challenges for software-defined networks." Communications Magazine, IEEE 51.7 (2013): 36-43.

\bibitem{Kreutz}
Kreutz, Diego, et al. "Software-defined networking: A comprehensive survey." proceedings of the IEEE 103.1 (2015): 14-76.

\bibitem{Scott}
Scott-Hayward, Sandra, Gemma O'Callaghan, and Sakir Sezer. "Sdn security: A survey." Future Networks and Services (SDN4FNS), 2013 IEEE SDN for. IEEE, 2013.

\bibitem{Kloti}
Klöti, Rowan, Vasileios Kotronis, and Paul Smith. "Openflow: A security analysis." Proc. Wkshp on Secure Network Protocols (NPSec). IEEE (2013).

\bibitem{Gong}
Gong, Xun, et al. "Website detection using remote traffic analysis." Privacy Enhancing Technologies. Springer Berlin Heidelberg, 2012.

\bibitem{Shin}
Shin, Seungwon, and Guofei Gu. "Attacking software-defined networks: A first feasibility study." Proceedings of the second ACM SIGCOMM workshop on Hot topics in software defined networking. ACM, 2013.

\end{thebibliography}

\end{document}